# A new empirical approach in the Search for Extraterrestrial Intelligence: Astrobiological nonlocality at the cosmological level

Fred H. Thaheld

**Abstract**

Over a period of several decades a concerted effort has been made to determine whether intelligent life exists outside of our solar system, known as the Search for Extraterrestrial Intelligence or SETI. This has been based primarily upon attempting to intercept possible radio transmissions at different frequencies with arrays of radio telescopes. In addition, astrophysical observations have also been undertaken to see if other worlds or solar systems exist with similar conditions such as ours which might be conducive to life. And, numerous papers have been written exploring different possibilities for the existence of life or why we have not observed it as of yet, since none of these approaches have been successful. It may now be possible to explore this issue from another standpoint. Recent theoretical and experimental results in the field of biophysics appear to indicate the possibility of quantum entanglement and nonlocality at the biological level, between spatially separated pairs of human subjects and also between basins containing neurons derived from human neural stem cells. If this research continues to be upheld in a more replicable fashion, this could have very important implications in the area of controllable superluminal communication. Experiments are proposed in an attempt to address the issue of whether controllable superluminal communication is possible and, if it is, to utilize it in an attempt to determine if extraterrestrial intelligence really exists, within the framework of astrobiological nonlocality.

*The only way of discovering the limits of the possible is to venture a little way past them into the impossible.*

                Arthur C. Clarke

**Introduction**

A research program has been going on for over 4 decades known as SETI, which consists mainly of monitoring selected sections of the sky with radio telescopes to see if one can pick up some type of narrow-bandwidth radio signal from space from another intelligent life source (Drake, Sobel, 1991; Zuckerman, Hart, 1995; Crawford, 2000). So far, despite massive worldwide distributed computer analysis, not a single signal has turned out to mean anything. The problem is that these signals can only move at the speed of light, which drastically reduces the range within which one can conduct a search, to just a few light years, as compared to a universe with an age of ~ 14 Gyr. And, due to the vast distances involved, any such electromagnetic signal would probably have become too weak long ago to detect on earth or, has become lost in the noise. The other major problem is that we would only be searching during a very limited period of time, known as the electromagnetic era for any advanced civilization, which probably encompasses a very short time period before they move on to more sophisticated techniques. The "communication window" (Ćirković, 2003) or "window of opportunity" then, for this type of search, is extremely limited as compared to the age of the universe and, when compared to other earth-like planets in the Galactic habitable zone alone, which are from 1.8 - 3 Gyr *older* than earth, with their implications for supercilivilizations.

If we were to pick up such a signal, it would have had to come from many light years away, meaning that the intelligent life which sent it has either disappeared or, more likely, has advanced technologically in the interim beyond the electromagnetic stage, to much more exotic communication techniques. If we attempt to respond, our signal would not reach them for a similar amount of time (if ever), making for a very cumbersome communication process where most of the participants would have long since expired.

Another problem is that if an advanced civilization were to encode and compress its electromagnetic communications, whether of a TV or radio signal nature, this would make these signals indistinguishable from the thermal radiation of the stars, and thus impossible to detect because it would seem like part of the universe's background noise (Lachmann et al, 2004). And, most likely the bulk of these advanced civilizations would have long ago progressed beyond the electromagnetic stage anyway.

A new scanning technique known as the Square Kilometer Array (SKA) will shortly be brought on line, able to probe further towards other possible earth-type civilizations (Penny, 2004). There is also the possibility of trying to detect macro-engineering projects over interstellar distances, which may prove to be more promising than searches for directed, intentional radio or microwave emissions (Ćirković, 2006).

Finally, with all the research that has been done to date regarding SETI, one has to raise the same question that Fermi did, "Where are they"? (Gato-Rivera, 2005). The scarcity of intelligent life in our Galaxy may have been due to gamma ray bursts which were lethal to land based life during the same period of evolution $\sim 10^8$ years ago (Annis, 1999). However, this same reasoning would be difficult to apply to all the other billions of Galaxies in the universe.

**Additional considerations**

I first started getting gradually interested in the SETI issue while doing some research for several different papers in the area of biophysics, and coming across early experiments which appeared to indicate brain wave correlations between pairs of human subjects (Duane, Behrendt, 1965; Grinberg-Zylberbaum et al, 1994), in addition to a provocative paper exploring the possibility of an interface between biology and quantum nonlocality (Josephson, Pallikari-Viras, 1991). Recent limited experiments of an increasingly sophisticated nature, continue to show the existence of these unusual brain wave correlations or autocorrelations between pairs of separated human subjects (Standish et al, 2004; Wackermann et al, 2003). Most of these were instances where one of the pair was subjected to some type of stimulus, such as patterned photostimulation, which results in the repeated appearance of a distinctive visual evoked potential (VEP) in the stimulated subject's electroencephalogram (EEG) at >8 µV. Correlated brain waves or anomalous events not possessing any wave-form similarity, appeared simultaneously in the non-stimulated subject's EEG at <2 µV, which event (even at the reduced µV levels) to say the least, is very counterintuitive! What lends further credibility to this research is that many of these subjects were in Faraday chambers, which screen out most of the possible electromagnetic influences or interferences of interest, in addition to any neural, acoustical, electrolytic or visual influences. Some of these experiments have also been conducted while the non-stimulated subject was undergoing functional Magnetic Resonance Imaging (fMRI), with similar results (Standish et al, 2003). The other unusual thing was that these experiments have been performed on several different continents, with implications for my proposal to be explored later. This appears to *imply* quantum entanglement and nonlocality at the biological level.

In addition, this same correlation effect has also been observed in a much more replicable fashion at the mV level between two separated and completely shielded 2 cm dia basins, containing neurons derived from human neural stem cells (Pizzi et al, 2004a;

2004b; Thaheld, 2000). I.e., these neurons represent two parts of a *common human DNA culture*, monitored by EEG. Laser stimulation of *one* neuronal basin containing neurons grown on a multielectrode array (MEA), not only caused an electrical response or action potentials from these neurons but, resulted in the appearance of *simultaneous* correlated electrical events at a high Hz rate in the second *non-stimulated* neuronal basin, while both basins were inside Faraday cages. Their initial experiments with a 670 nm laser and just *one* MEA, showed very high values of crosscorrelation or autocorrelation and frequency coherence during the series of laser impulses, between the laser and this *single* stimulated basin (Pizzi et al, 2006). The important thing here is that this repeated laser stimulation and the resulting simultaneous electrical responses from *both* the stimulated and non-stimulated basins, *implies* that quantum entanglement can be maintained in biological systems, and resists or uses decoherence in a *constructive* rather than the usual *destructive* fashion, thereby maintaining biological entanglement laser stimulus cycle after cycle (Matsuno, 1999; 2006; Thaheld, 2003; 2005a).

**Further analysis and basic assumptions**

I became further interested in the SETI problem while exploring a new approach to the measurement problem (Thaheld, 2005b; 2006a). After coming to the conclusion that wave function collapse might take place in a natural fashion in the rhodopsin molecule in the retina, in an objective biophysical manner (a position previously taken by Shimony, 1998), it then occurred to me that if this were true, then Everett's 'many worlds' or 'many minds' theory (Everett, 1954), which he had proposed as an alternate explanation in lieu of wave function collapse, had to be incorrect. The next logical step was to apply the 'many minds' concept to ongoing SETI research but, within a possible controllable superluminal communication (CSC) framework (Thaheld, 2006c).

In order to set the stage for the following proposal, it is necessary to make some basic assumptions and, to briefly recap some of the prior biophysical theory and research upon which this approach is based.

First, I assume that extraterrestrial intelligence does exist, just like the present practitioners who have been attempting to detect radio waves for several decades. And, that it consists of the same basic components necessary for life and replication as we do, as regards DNA and RNA, having arisen from the same universally available primordial constituents that were responsible for the appearance of life on our planet.

Second, that most of this extraterrestrial intelligence is superior to us (probably by vast levels) or otherwise it will not be capable of sending, receiving and interpreting the type of nonlocal information transmission being proposed in this theory. A whole host of other life forms inferior to us in varying degrees and in different stages of the evolutionary cycle, must also exist throughout the universe, capable of unknowingly and unconsciously sending out random nonlocal signals but, probably not in a signal-recognizable fashion and, in more of a chaotic nature.

Third, that there must be some type of action potential involved in this process and, that without its presence, there is probably not going to be a possibility to achieve the CSC being proposed. By way of explanation, an action potential is a result of a neuron firing, going from 'off' to 'on' in a rapid fashion as usually measured in ms and, with a large swing in voltage amplitude from a resting level of $\sim -70$ mV to $\sim +10$ mV during the firing process. I have previously touched upon the importance of the action potential and the role it must play with regards to biological entanglement and nonlocality (Thaheld, 2005a), and also more recently with regards to CSC (Thaheld, 2006c). That in effect, when a neuron fires, one gets both a classical and nonlocal output or *signal* and, that it is this nonlocal *signal* or *information*, that can be received by another neuron or neurons at any distance in the universe, and converted back into a classical signal, capable of being deciphered or read. This type of nonlocal *information* does not conflict with special relativity. This could also be considered to fall into the category of the 'reverse direction' problem, which I attempted to deal with in a prior paper (Thaheld, 2003). I.e., we know that neural events can initiate or cause mental events so, it must also be possible for the 'reverse' to happen, that mental events can initiate or cause neural events, which is important in this proposal.

Fourth, that if biological quantum entanglement and nonlocality exist at both the local and cosmological levels, this would then make it possible to achieve CSC. Without CSC *instantaneously* over any distance in the universe, the present proposal becomes impossible to achieve. Also, it is important that this biological entanglement has to be capable of being maintained or regenerated, no matter what distance is involved, either in opposition to decoherence or, feeding off of decoherence in some constructive manner, measurement after measurement (Matsuno, 2006; Thaheld, 2005a). This ability appears to have been achieved in most of the experiments previously cited, not only in the case of the human subjects but, especially in the case of the basins containing the human neurons, due to the continuous Hz stimulation by the laser over varying periods of time. This would not have been possible if entanglement had been broken.

Fifth, that we are constantly being bombarded by this presently undetected type of superluminal communication, and have been for a very long and unknown time. That the universe is awash in it, just as it is with neutrinos, cosmic rays, the cosmic microwave background, gravitational radiation and the entire electromagnetic spectrum. Superluminal signals are constantly arriving from advanced and not so advanced civilizations, with varying degrees of complexity, probably looking like a lot of noise (one only has to look at an EEG record to see what just human brain wave EEG noise looks like) and, that what we looking for is probably within this 'noise'. It will just require a special but, very simple means of detection (which is already readily available), just like the special equipment and techniques which is required in the search for neutrinos, cosmic rays and the ongoing search for gravitational radiation. The EEG data gathered will also have to be run through a specially designed series of algorithms.

**The most critical element: The universality of the ingredients for life**

The origin of life as a planetary phenomenon will probably resist successful explanation as long as we lack an early record of its evolution and additional examples (Gaidos, Selsis, 2006). Plausible scenarios for the origin of important prebiotic molecules and their polymers on the earth, involving atmospheric chemistry, meteorites, deep-sea hot springs (Matsuno, 2006) and tidal flat sediments have been developed. Terrestrial life may not have originated on the earth or even on any planet. The smaller primitive bodies of meteorites, in which carbon molecules and catalytic transition metals were abundant, and in which hydrothermal circulation persisted for millions of years, offer alternative environments for the origin of life in our solar system (Gaidos, Selsis, 2006).

The building blocks of proteins, amino acids, are the essential molecular components of living organisms on earth, and researchers are confident that they exist in planetary systems throughout the universe (Bernstein et al, 2002; Caro et al, 2002). This means that all of the hospitable planets throughout the universe would have probably been seeded with specific forms of amino acids, which are molecules that normally come in mirror-image right- and left-handed forms, and form the proteins used by life on earth (Bernstein et al, 2002; Caro et al, 2002). All naturally occurring proteins in organisms on earth use the left-handed forms, which raises the question as to how and when this chirality came into play. Meierhenrich (2005) feels that the right-handed molecular building blocks of life, the amino acids, were preferentially destroyed by circularly polarized light. That these amino acids were created in interstellar space and came to the earth in the form of comets or micrometeorites. He does not rule out that other solar systems may harbor right-handed amino acids, if they have been subjected to the other type of circularly polarized light.

Natural proteins are strings of 20 different amino acids which can be made by simple chemical synthesis in water, while meteorites containing more than 70 types of amino acids, have been bombarding earth from its very beginning (Bernstein et al, 2002). This means that *biomolecules are universal*, which raises the potential for the emergence and evolution of life throughout the universe, which may then be expected to possess many critical similar characteristics as are found here on earth, with special emphasis on intelligence. The success of this proposal hinges upon certain critical similarities. Extraterrestrial beings would have probably resembled us at the same stage in evolution, which would not be the case now due to the passage of vast amounts of time, which of course implies the evolution of many unusual design changes of a vastly improved nature, especially as regards an advanced level of intelligence with commensurate communication capabilities.

These proteins and amino acids play an essential role in the creation of DNA and RNA. DNA contains all of the genetic information necessary to construct cells, to integrate them into an organism and to maintain them (Brill, 1995). RNA translates this information into specific instructions for the assembly of proteins, transmits the information outside the cell nucleus and helps to assemble them. DNA is found mainly in the nucleus of the cell. RNA is found in the cytoplasm of the cell but, is synthesized in the nucleus. DNA contains the genetic codes to make RNA, RNA then contains codes to make proteins. Proteins are the chemicals of life. The proteins and nucleotide sequence are slightly different for each individual. The types of proteins and the length of DNA are different for each species. But, we share with all other people and all other forms of life the same set of amino acids, the same nucleotides and the same genetic code.

**Applicable theory and research**

The question of nonlocality was first raised by Einstein-Podolsky-Rosen (EPR), who claimed that if quantum mechanics were a complete model of reality, then nonlocal interactions between particles had to exist (Einstein et al, 1935). Bell later directly addressed this problem of nonlocality, and proposed an experiment to test for its validity (Bell, 1964). An experiment was later performed that showed that nonlocal influences do exist once these particles interact and, that one can test the explicit quantum nature of systems via the use of EPR nonlocality (Aspect et al, 1982). And, as per Feynman, since this nonlocality cannot be duplicated by a classical system, this enables it to be used to test the quantum nature of systems (Feynman, 1982). Of course, Einstein, Bell, Aspect and Feynman, among many other physicists, always considered this issue of nonlocality to be applicable only to *inanimate* objects such as particles and, never considered that this same process could be applied to *animate* entities in the biological realm. Their reasoning appears to have been incorrect as has been demonstrated by later research as illustrated by the following.

Over 40 years ago the first experiment was performed which appeared to reveal rather highly unusual and unexpected correlations between the brain waves or EEGs recorded from pairs of spatially separated identical or monozygotic twins (Duane, Behrendt, 1965; Thaheld, 2001). The researchers had noted that the nonscientific literature was replete with instances in which illness or trauma in one of a pair of identical twins affects the other, even when they are *far apart*. They decided to alter the brain wave pattern of one twin and see if this would produce a similar response in the brainwaves of the other twin. In this instance, the alpha rhythm was utilized, which are brainwaves of from 8-13 Hz and <50 µV. This alpha rhythm can be elicited when the subject closes his eyes, stares at a uniform unpatterned background or, when he sits in the dark with his eyes opened. When one of the twins was asked to close his eyes, it was noticed that the alpha rhythm not only appeared in his EEG, but also in the EEG of the other twin who had kept his eyes open. This is highly unusual to see an alpha rhythm in this twin, as one normally expects to see a beta rhythm consisting of 14-25 Hz brainwaves at <20 µV. This effect was observed in only 2 out of 15 pairs tested and, while the sample for statistical purposes is admittedly low, one does not expect to see anything of this unusual nature! And, if one keeps repeating this experiment with the same results, it becomes as statistically significant as any physics experiments performed with low efficiency detectors concerning the Bell inequalities (Aspect et al, 1982).

A more detailed analysis of numerous EEG studies performed over 6 decades, reveals that a majority of identical twins have very similar EEGs and mental habits, whether within or outside the normal range (Thaheld, 2000; 2001; 2005a). This similarity of EEGs by itself, should not be construed as evidence for quantum entanglement, as phase coherence also occurs normally in the classical world without entanglement.

Later studies continued to show correlations between the EEGs of stimulated and non-stimulated subjects, leading up to a pioneering study with a larger number of subjects (Grinberg-Zylberbaum et al, 1994), in which one of a pair was subjected to visual stimuli (light flashes from a Grass photostimulator), while the other subject was not being stimulated. It was noted that some of the non-stimulated subject's brain wave forms, as reflected in their EEG record, were *simultaneously correlated* with the stimulated subjects VEP-elicited brain wave forms, which normally result from the visual stimuli. The non-stimulated subject's brain waves were not exact replicas of the stimulated subject's waveforms nor were they of the same amplitude. Once again it is highly unusual for one to see *any correlated brain wave events* from the non-stimulated subject, since both subjects were in Faraday chambers, which effectively screens out most of the electromagnetic waves of interest, and any neural, acoustic or visual effects.

The problem which researchers in this field face is that although the elicited VEP in the stimulated subject's EEG carries a distinct amplitude and frequency, it is in a very low microvoltage range to begin with and, the microvoltage in the non-stimulated subject's EEG is not only several orders of magnitude lower but, his wave-form, as regards to both frequency and amplitude, comes nowhere near to being a copy of the original initiating VEP. That the frequency signature is not retained in the VEP of the non-stimulated subject represents a serious defect analogous to low detector efficiencies.

Later, more sophisticated experiments were conducted in which patterned or checkerboard photostimulation was used to investigate possible EEG correlations between human brains (Wackermann et al, 2003: Standish et al, 2004). While the experimenters at the Univ. of Freiburg have achieved a more robust replication of the Grinberg-Zylberbaum research, they have noticed a similar problem, in that while the *averaged* stimulated subject's VEP peak amplitude is >8 µV, the *averaged* non-stimulated subject's correlated VEP is <2 µV, with no wave-form similarity between the initiating VEP and the resulting correlated VEP of the non-stimulated subject (Wackermann et al, 2003-See Fig. 1). Their results indicate a high co-incidence of variations or correlations of the brain electrical activity in the non-stimulated subjects with brain electrical responses of the stimulated subjects. They did not see any VEP-like wave-forms in the averaged EEG of the non-stimulated subjects.

Experiments have also been performed where one of the pair was undergoing a flickering checkerboard pattern stimulation, while the non-stimulated subject was undergoing an fMRI (Standish et al, 2003). Changes in fMRI brain activation (relating to blood oxygenation) and EEG signals during the stimulus condition, were observed in the non-stimulated subjects. Even with a very small subject sample base (especially in the case of fMRI), these experiments have continued to corroborate, with increasing experimental and statistical sophistication, these unusual EEG and fMRI correlations, which appear to have no classical explanation.

As has been previously mentioned, research has been ongoing for over 3 years at the University of Milan (Pizzi et al, 2004a; 2004b; 2006; Thaheld, 2000), utilizing pairs of 2 cm dia basins containing neurons derived from human neural stem cells attached to MEAs inside Faraday cages, separated by over 20 cm. The voltages in these neuronal basins, prior to laser stimulation, are ~5 mV peak to peak. Laser stimulation of just *one* of the basins at 670 nm, results in an electrical signal being generated by this basin, which can vary from 0-2,000 Hz and, with a peak to peak amplitude of 20 mV. There is a normal delay between the time of activation of the laser, its impingement upon the neurons and the resulting electrical signal, of some 300 ms. What they have found most interesting over a period of several years and thousands of laser pulses, is that the *separated non-stimulated* basin displays simultaneous or correlated electrical signals, of a similar amplitude in mV and, that there is a simultaneity or correlation in the 2 basin's frequencies between 500-2,000 Hz, with a sharp common peak around 900 Hz. Periodograms of the two signals are also about coincident.

They have resorted to every possible technique (as detailed in their papers) to rule out the possibility that this effect might be due to some error in their experimental protocol, in equipment malfunction or some type of human intervention, since the 2 neuronal basins are only separated by a few cm. To show you the extent they have gone to, to rule out any possible local or classical explanation, they first start out with just one neuronal basin and 2 control basins without any neurons in them but, containing either their culture liquid or matrigel. Both basins are inside Faraday cages. They stimulate the main basin containing the human neurons on the MEA with the laser, and get the usual electrical response from it but, there are never any electrical signals or response from either of the two control basins.

Their research reveals consistent electrical waveform correlations between the stimulated and non-stimulated basins, on a much more repeatable and replicable basis than any of the human subject research performed to date, with a considerably better signal-to-noise ratio, and with amplitudes as measured in the mV range, as compared with the poor signal-to-noise ratio and the very low µV coming from the brain. It is very critical to note here that repeated stimulation at varying Hz rates with either patterned photostimulation with the human subjects or with laser stimulation of the neuronal basins, does not break this entanglement, as is especially evident in the case of the neuronal basins. This indicates that after every 'measurement', quantum entanglement is either regenerated or maintained in some fashion, contrary to expectations (Tegmark, 2000; Hagan et al, 2002; Thaheld, 2005a). There are also additional indications that biological quantum nonlocality has been observed in the coherence of induced magnetic dipoles involved in muscle contraction in single actin filaments at the mesoscopic level (Matsuno, 1999; 2001). All of this evidence, when taken together, indicates that we could be looking at a generic phenomenon.

One of the main drawbacks in this field of endeavor, especially with human subjects, is the lack of a large sampling base upon which to base more solid statistics. However, it should be pointed out that even with this small human test sample base, one is constantly stimulating at varying Hz, so that over a short period of time one can get thousands of discrete events revealing themselves on the EEG or in an fMRI, which can then be subjected to averaging techniques. I have also attempted to address the other main drawback in this field of trying to find more reliable and replicable types of stimuli, including the use of transcranial magnetic stimulation (TMS), where several areas of the brain are *simultaneously activated or deactivated* as a result of this stimulation (Thaheld, 2001; 2003; 2006b). In addition, this can also be correlated with a measured external response such as an evoked motor potential in a particular muscle group such as the hand or leg, or by the appearance of phosphenes in the occipital cortex. It may also be possible in the near future to examine whether other animals such as chimpanzees or dolphins, possess this quality of biological quantum entanglement, since they preceded us and we all share the same genetic code (Thaheld, 2004).

**How does one achieve entanglement at the biological level not only in our world but throughout the universe?**

In one of my papers (Thaheld, 2003), I had noted that biological entanglement must already exist throughout the world, since experiments conducted to date over several decades and on several different continents, repeatedly show electrical correlations or autocorrelations between the brain waves of not only related but unrelated human subjects of different nationalities, indicating that this is a generic phenomenon. While the sampling for statistical purposes is extremely sparse, it is still significant that certain anomalous effects continue to not only appear at various Hz stimulation levels but, to be maintained in the face of what should be destructive decoherence (Tegmark, 2000; Hagan et al, 2002). The question that I next addressed had to do with just how this generic phenomenon might arise within a biological setting (Thaheld, 2003). The simplest answer seems to be in the fact that all humans on the face of the earth have similar features in their DNA, as has been revealed from blood samples taken from different nationalities around the world. This suggests that we all have a common human origin going back to a single ancestor ~ 170,000 years ago.

This would appear to meet the requirement laid down by Schrödinger, that in order to achieve entanglement, objects must have been in contact with each other (Schrödinger, 1935). This is certainly the case for DNA, which was not only in contact originally but, repeated and maintained this critical process of contact through the endless cycles of replication. It also occurred to me that this same process might be going on at all levels of the animal world, so that we really have to go back beyond the common human ancestor approach, since we are descendents from them and, here once again, share a lot of similar DNA (Thaheld, 2003). While this is not necessarily pertinent to the present paper, the author has devised several experiments to test this theory, involving either pairs of chimpanzees or dolphins initially, and proceeding down the evolutionary ladder, essentially adopting the same experimental techniques which have been used for human subjects (Thaheld, 2004).

I have also explored the critical issue of the difference of entanglement and nonlocality between *inanimate* objects such as particles and *animate* or biological objects, and why this entanglement could be both maintained and transferred in the case of biological systems, even in the face of decoherence (Thaheld, 2005). This entanglement persists even in the face of repeated measurements at varying Hz rates, which is important for the present proposal.

What this could imply is that we may then be related to the first elemental life which sprang forth on this planet over 3 Gyr ago, and that all living entities are entangled in quantum mechanical fashion. This can also be subjected to fairly simple experiments as outlined.

The thought has now undoubtedly occurred to you as to how one could even possibly go about achieving entanglement at the cosmological level, and still adhere to Schrödinger's dictum regarding the necessity for prior contact, especially between objects spatially separated by vast light-year distances?  There appear to be several approaches which one might take to this problem, in addition to what might be dreamt up by worm-hole, string, brane, multi-dimensional and multi-universe theorists.  To properly set the stage it will be helpful to quote Schrödinger on this subject of entanglement (Schrödinger, 1935).

"When two systems, of which we know the states by their respective representatives, enter into temporary physical interaction due to known forces between them, and when after a time of mutual influence the systems separate again, then they can no longer be described in the same way as before, viz. by endowing each of them with a representative of its own.  I would not call that *one* but rather *the* characteristic trait of quantum mechanics, the one that forces its entire departure from classical lines of thought.  By the interaction the two representatives (or wave-functions) have become *entangled*".  We are now in a better position to realistically discuss the approaches which one might take to this problem.

First, if one is an adherent of the Big Bang theory, then our entire universe was, at the very instant of creation, encompassed within a singularity, which would represent the ultimate concept of contact, exceeding by far the most massive black holes and appearing to more than meet Schrödinger's dictum.  This would imply some type of universal entanglement at the most fundamental level for all particles, and the later possibility of either maintenance of the same original levels of entanglement as the universe expanded and developed or, the creation of different sub-levels of entanglement when life commenced.

Second, it is not beyond the realm of possibility that any sufficiently advanced or supercivilization could have developed techniques of generating and maintaining entanglement between any living entities in the universe in a nonlocal, instantaneous fashion. Or, gone beyond the concept of CSC.

Third, to paraphrase Dirac, this problem is not ripe for solution at the present time and should be left for later (Dirac, 1963). I.e., not to worry, that it will be solved at a later date!

**The issue of controllable superluminal communication: Is it attainable within the framework of biological nonlocality?**

With a universe now some 14 Gyr old, just how is it possible for there to be any type of CSC between any vastly separated worlds on which there might be superior intelligent life? After all, this matter of controllable vs uncontrollable superluminal communication has generated a lot of controversy in the scientific community. While it has been shown through experiments that superluminal effects do indeed exist between entangled particles (Aspect et al, 1982; Zbinden et al, 2001), it is generally thought that this effect cannot be used to achieve CSC (Shimony, 1984).

Relativity theory postulates the non-existence of faster-than-light 'signals' but, does not necessarily impose an analogous requirement upon all other conceivable kinds of 'influences' (Stapp, 1988). He proposes the existence of superluminal 'influences' between the entangled photons which are not considered to be 'signals', with the result that no conflict with the theory of relativity is entailed. Based upon what is known as the Eberhard theorem, it is felt that no information can be transferred via quantum nonlocality (Eberhard, 1978). Shimony has also come to the same conclusion that the nonlocality of quantum mechanics is uncontrollable, and cannot be used for the purposes of sending a signal faster than light (Shimony, 1984).

Why is it that the entangled photons in the Aspect experiment cannot be used to transfer information faster than light, keeping in mind that these are *inanimate* entities when commencing such an analysis? After all, it has been shown that nonlocal correlations exist between these photons and one would logically think that you should be able to perform this feat. The polarization correlations cannot be used to transmit information faster than light because they can only be detected when the statistics from the measurements on each side are compared in a classical fashion, which is dependent upon the efficiency of the detectors. The act of polarization measurement on photon v1 forces it to move from the quantum indeterminate level to a specific determinate level, and it is this information that is transmitted nonlocally to entangled photon v2. When Alice, let us say, measures the polarization of photon v1 along a direction she chooses, she cannot choose the result nor she predict what it will be. Once she makes her measurement, Bob's photon v2 simultaneously receives nonlocal information regarding a similar state of polarization, where he cannot choose the result either. Since Alice has no control over the results she gets, she cannot send any meaningful information of her own to Bob. Similarly, Bob can choose one of several polarization measurements to make but, he will not know the result ahead of time. Alice and Bob can only see the coincidence of their results after comparing them using a conventional method of communication, which does not send information faster than light.

I feel that while this reasoning is correct for *inanimate* or *nonliving* entities such as photons where, once a measurement has been made, the wave function collapses and they become disentangled, the situation is much different for *entangled living* entities such as the human subjects and the human neurons on the MEAs. Here, instead of being limited to just a pair of entangled photons at the microscopic level, which only briefly spring into existence and then are gone, we are looking at a massive number of *living* entangled particles at the macroscopic level, either resisting the usual decoherence or utilizing it as has been postulated (Matsuno, 2006; Thaheld, 2005) to either maintain or regenerate entanglement after each measurement, thereby enabling us to achieve CSC (Thaheld, 2000; 2001; 2003).

**The proposed experiments**

The proposed experiments can take several different directions, all of them fairly simple and inexpensive, as compared to present SETI experiments, utilizing existing electroencephalographic techniques and relying upon a much simpler computer-assisted analysis which runs the data through a series of algorithms designed to detect signals that have some possibility of being both artificial and either of a terrestrial or extraterrestrial nature. We can directly test for the actual existence of CSC here on earth either through the use of pairs of human subjects or neurons derived from human neural stem cells attached to pairs of MEAs. The present researchers may have already been observing nonlocality, which implies superluminal communication in their experiments but, they have been hesitant to come right out and say this due to the obvious risks involved if one makes such a startling statement based upon limited experimental data, and it turns out to be incorrect.

The research group at the Univ. of Freiburg, dealing with the human subjects, stresses that "while no biophysical mechanism is presently known that could be responsible for the observed correlations between EEGs of two separated subjects, nothing in our results substantiates the hypothesis (Grinberg-Zylberbaum et al, 1994) of a direct quantum physical origin of correlations between EEGs of separated subjects" (Wackermann et al, 2003). The researchers at the Univ. of Milan state that, "despite at this level of understanding, it is impossible to tell if the origin of this nonlocality is a genuine quantum effect, our experimental data seem to strongly suggest that biological systems present nonlocal properties not explainable by classical models (Pizzi et al, 2003).

There are 4 major problems which must be addressed. First, the number of human subjects must be increased, along with the number of neuronal basins. Second, it will be necessary to adopt the use of different types of stimuli, in addition to patterned photostimulation, such as TMS, so that we get more replicable and discernible results from several different areas of the brain simultaneously (Thaheld, 2001; 2003, 2006). And, in addition to the black and white patterned or checkerboard photostimulation, we may have to resort to various mathematical or other type of stimulus symbols as a better means of exploring the mind-brain interaction (Thaheld, 2003). Third, it will be necessary to separate the pairs of human subjects beyond the present several meters, to much greater distances to determine if the repetitive anomalous effects are still observed without any diminution in the non-stimulated EEGs or fMRIs. This also applies to the pairs of neuronal basins, where the distance of separation is now measured from several cm up to a meter or more. Fourth, while the stimulation of just one entity (whether human subjects or MEAs), which presently results in the generation of electrical signals and, the simultaneous reception of a correlated signal (not necessarily of similar Hz or amplitude characteristics by non-stimulated entities), is very indicative of quantum nonlocality, this hypothesis can be significantly strengthened if we reverse the situation. I.e., the non-stimulated entity now becomes the stimulated one and vice versa. If this effect is achieved, this would be direct evidence of CSC at a very rudimentary level, since both entities would be exchanging signals at the same variable Hz rate and confirming same via a non-classical means, since they are all inside Faraday chambers.

This would be directly related to prior research (Grinberg-Zylberbaum et al, 1994) that since photostimulation was utilized to elicit a VEP in the brain waves of the stimulated subject, that when a flickering light signal is used, the normal VEP often carries a frequency signature. One of the researchers (Goswami) has proposed that to the extent that this frequency signature is also retained in the transferred potential elicited in the non-stimulated subject's brain, it may be possible to send a message, at least in principle, using a Morse code by varying the frequency of photostimulation to resemble a code. He has further suggested that the brain obeys a nonlinear Schrödinger equation in order to include self-reference (Mitchell, Goswami, 1992) and, that for such systems, message transfer via EPR correlation is permissible (Polchinski, 1991). This directly ties into a recent proposal to resolve the measurement problem (Thaheld, 2005b; 2006), with collapse of the wave function taking place within the retina of the eye in a nonlinear objective fashion, either as a result of conformational change (s) in the mesoscopic retinal-rhodopsin molecules or, as a result of the electrical amplification process immediately following these changes. An alternate analysis of the physiology of the eye and reduction of the state vector from a Continuous Spontaneous Localization (CSL) standpoint, reveals that reduction of the state vector may take place directly within the individual rods of the retina during the amplification chain in a rod cell (Adler, 2006).

Whichever reduction process turns out to be correct, since they both call for a collapse of the wave function, this means that the Schrödinger equation has to be modified to reflect this nonlinear process, lending further support to a CSC hypothesis.

**1. Human neurons on MEAs**

The simplest way to check out whether entanglement and CSC are possible in a nonlocal fashion here on earth, is to use 2 basins containing human neurons mounted on MEAs (Thaheld, 2005a; 2006c). The 2 basins would be separated by several meters with both inside Faraday cages. Just one basin is stimulated repeatedly at a random Hz rate with a laser, resulting in an electrical response from the stimulated neurons which is correlated with the stimulus rate. If past research is any criteria, this same Hz rate should also be observed in the 2$^{nd}$ non-stimulated neuronal basin. This would represent preliminary evidence for both entanglement and nonlocality. In order to achieve CSC, it is only necessary to reverse this situation, with a laser now stimulating the formerly non-stimulated basin, and repeating the same random Hz rate, returning the same message back to the formerly stimulated basin. The 2$^{nd}$ basin has, in effect, *informed* the 1$^{st}$ basin that it has received its initial message by sending the message back to it without the use of any classical means of communication!

The natural objection which one can make to this experiment is that the distance of separation between the 2 basins, being only a meter or less, does not rule out a local or classical explanation, that the Faraday cages are not effectively screening out all electromagnetic influences of interest or, that the experiment is flawed in some unknown fashion. Since these types of experiments with the neuronal basins have been conducted through thousands of cycles over a period of several years, with increasingly stricter standards and the same positive results, these possibilities would appear to be remote.

The way to determine if entanglement and nonlocality really exist is to modify a technique used by the Italian researchers (Thaheld, 2006c). In their experiments involving learning in human neural networks on *single* MEAs (Pizzi et al, 2006), they stop these neuron's electrical activity or action potentials, to insure that the collected signals are actually due to the electrophysiological functioning of these neurons. They are able to accomplish this by injecting the neuronal culture on the MEA with Tetrodotoxin (TTX), which is a neurotoxin able to abolish action potentials. They once again stimulate this TTX treated neuronal basin with the laser and get no electrical response, showing the importance of the action potentials, or rather the pores of the voltage-gated sodium channels in nerve cell membranes, since TTX binds to these pores, thereby blocking the action potentials in the nerves. The TTX is then rinsed away, the action potentials return, they stimulate with the laser once again and note that the previously correlated electrical signals have returned.

In the modified experiment we once again use 2 separated MEAs and, after making sure that they are entangled by stimulating one with a laser and getting an electrical response from both the stimulated and non-stimulated basins, inject the non-stimulated basin with TTX, thereby stopping all action potentials. We then stimulate the untreated basin with the laser, and should observe the usual electrical signals being generated by this basin but, with no corresponding correlated electrical signals coming from the non-stimulated TTX treated basin. To double check, we next stimulate the TTX treated basin with the laser and we should not only get no electrical signals from it, but also no electrical signals from the untreated basin! If this is what happens, we have immediately proven three things: First, there are no exterior influences or experimental design flaws which could account for these unusual correlations. Second, that the action potential plays a very important role in this process and, that if it is in any way impaired or stopped, the entanglement and nonlocality between the two MEAs comes to a halt. Third, that CSC is possible!

The second experiment would be conducted with the SETI program in mind. This would involve subjecting just one neuronal basin of a pair to a random series of stimulation from the laser. We would then leave this whole system in isolation for a period of hours, during which time we are recording any resulting electrical activity from this basin and its companion non-stimulated basin. Later, we run the data through a series of algorithms for a computer-assisted search for certain signals which might have been received or generated by these neurons, which can either be of a similar repetitive nature to the original random Hz rate or, are of a more interesting random or repetitive nature, either as regards Hz rate or amplitude. I.e., it is something which you would not normally expect to see. Now the question becomes whether these signals could have originated from some source here on earth or that they truly represent some extraterrestrial origin. It is highly unlikely that someone somewhere on earth could somehow be using this same random stimulus sequence in a similar experiment. The only thing we would all have to agree upon in advance is what type of random stimulus signals we are going to use, and the Hz and amplitude range that we want the computers to search within, of either a random or unusual nature.

### 2. Human subjects

The next step is to determine if entanglement, nonlocality and CSC might exist between human subjects and, if it does, how this can be utilized with regards to the SETI matter (Thaheld, 2005). Two subjects, let us call them Alice and Bob, are placed in individual Faraday cages separated by several meters, with both of them hooked up to individual EEGs and with separate means of stimulation. Alice is subjected to either patterned photostimulation or to TMS, and a VEP is recorded on her EEG of a distinctive Hz and amplitude level. We then want to determine if the non-stimulated Bob elicits or receives a simultaneous event on his EEG which, while it will not necessarily resemble either the Hz or amplitude of Alice's EEG, will still represent a repeatable simultaneous anomalous event of a type which one would never observe under normal conditions but, which is correlated with the stimulation of Alice.

We then reverse this situation and now stimulate Bob and see if we still observe this same sequence of events in the case of Alice. If we do, we now administer a general anesthetic to Alice and, after the loss of consciousness, we subject Bob to the same stimulus as before and a VEP is recorded on his EEG. We now want to see if a simultaneous anomalous event is still recorded on Alice's EEG as it was before. If it isn't, this will show that entanglement and nonlocality was broken by the administration of the anesthetic, just as it happened between the MEA neuronal basins. To make sure we stimulate the anesthetized Alice with a flash stimulus and, even though we get a VEP on her EEG, we should get no anomalous response on Bob's EEG (Nuwer, 1986; Moller, 1988).

We now allow Alice to recover from the anesthetic, stimulate her once again, and see when Alice elicits a VEP this time, if simultaneous anomalous events again appear on Bob's EEG. If they do, this would mean that entanglement was not lost as a result of the anesthesia. Once again, we reverse this situation and stimulate Bob and see if simultaneous events appear on Alice's EEG.

Another technique which could be used is to have both Alice and Bob in Faraday cages with Bob asleep. Normally when a person is sleeping, you will observe very slow delta waves of less than 4 Hz with an amplitude ranging up to 100 µV. We can now stimulate Alice, who is awake displaying alpha or beta waves, and see if Bob's delta waves simultaneously change to either alpha or beta or display anomalous events tied in to the specific type of stimulation used on Alice. If we observe such an event, this would once again indicate entanglement between them. We are now ready to proceed to the SETI phase of this experiment.

It should be first noted that humans possess an assortment of different types of brain wave frequencies and accompanying amplitudes within a normal range as shown below:

| | | |
|---|---|---|
| alpha | 9-14 Hz | <50 µV |
| beta | 15-40 Hz | <30 µV |
| delta | 0.1-4 Hz | ~100 µV |
| gamma | 26-70 Hz | ~40 µV |
| theta | 5-8 Hz | ~10 µV |

There are also brain waves associated with epileptic seizures referred to as petit mal ranging from 2-4 Hz and with amplitudes up to 1 mV, and grand mal at >20 Hz and >500 µV. There are also electrical phenomena known as 'spikes' and 'saw-tooth' or 'square waves'. In addition, there occurs a wide spectrum of waves ranging from 100-1,500 Hz and upwards (Niedermeyer, E., 2005). For a very detailed treatment of this complex EEG brain wave issue please refer to (Niedermeyer, Lopes da Silva, 2005). It is of interest to note here that the true frequency range of the EEG is much broader than it has been assumed and taught for decades (Niedermeyer, 2005). With the introduction of digital EEG machines, the exploration and ultra fast recording of the 60 to 1000 Hz range has already begun in the past few years (Rodin, 2005). Ultra fast EEG activity has also been recorded at 3000 Hz.

It should be emphasized here that digital technology, when properly utilized, allows for a much better assessment of the informational content of EEGs (Rodin et al, 2006). The use of appropriate filter settings as regards Hz, and viewing windows as regards time, will be two of the critical elements enabling us to pinpoint any wave characteristics of interest, along with both compression and expansion of data, which also have their individual advantages. As an example, in present clinical EEG practice, unless filters are adjusted for a specific clinical question, one may miss critical information. The present software installed on digital EEGs may not prove adequate for the SETI task at hand and, even that which has been recently developed for further, advanced evaluation of digital EEGs may have to be supplemented by specifically tailored algorithms.

It has been noticed by EEG neurophysiologists and clinicians that an EEG filtered for a certain Hz range, may appear normal in channel tracings but, if the same data is looked at from a top view and different filter settings, this may reveal (let us say) delta activity which has been masked by alpha waves. Data is considerably easier to interpret when top views can be examined rather than just the conventional channel arrangement. Even with increased amplifications, widespread fast activity which was previously missed, becomes apparent.

It is also important to mention the frequencies below the delta range or subdelta, and the argument for what is referred to as "Full-band EEG", which ranges from DC to above 1000 Hz (Vanhatalo et al, 2005). Subdelta activity (0.5-<0.1 Hz) is currently being recorded by digital EEG instruments, without clinicians being fully aware of the fact. It is possible to view an ultra slow 0.01 Hz event with a wave duration of 100 seconds, provided that one has a long enough viewing window. With filter settings of 0.01-0.1 Hz it is no longer appropriate to speak of frequencies but, rather of events with various wave lengths. Different conclusions can be reached depending upon what filter settings and which time windows are used. In the future instrument manufacturers will have to incorporate in their software programs the opportunity for not only the typical clinical-type user but, for any SETI-type user to be able to set the desired filters or frequency bands and viewing windows, rather than only providing the present limits. This would then allow any clinician, technician or neurophysiologist who so chooses, to expand his usual diagnostic procedures, with very little additional effort, into a SETI-type search with each consenting patient!

I have only dealt very briefly with what is a very difficult area for doctors, neurophysiologists and technicians, even when they are attempting to diagnose subjects in their immediate vicinity, to show you how hard it will be to broaden the use of EEG instruments into any kind of even minimal and realistic SETI-type research.

Let us now proceed to the proposed human SETI experiment. We will essentially use the same technique as was proposed for the MEAs. We first stimulate one or more subjects with a random patterned or TMS stimulus, such that the likelihood of seeing anything like this occur on these subjects' EEGs is extremely remote. I.e., no one on the face of this earth would ever be expected to accidentally duplicate this pattern. We now want to record these subject's EEGs over a period of days and under all conditions of sleeping and waking via ambulatory means (Gilliam et al, 1999). Ambulatory recording or AEEG can be performed for up to 3-5 days without the need to change the batteries or the memory card, based on 19 channels of data recorded at 200 Hz. With an 'on the fly' battery and memory change, one can record indefinitely, with minor interruptions. We can have event markers set to determine if anything remotely approaching the initial stimuli configurations appears on the recorded EEG data stream, and can also use computer-assisted analysis to check for anything of an unusual or repetitive Hz or amplitude nature which is never seen in the normal subject's EEG. We would simply have to agree in advance as to what would constitute an event or events on the EEG which might be attributed to a SETI input, or within what ranges we want to establish a series of algorithms to run the data through for computer-assisted analysis, so that we rule out any possible artifact arising from any other sources on earth. This matter of agreeing in advance as to the critical boundaries we want to establish for any series of algorithms through which the data will be run, should prove to be the most difficult part of this proposal, just as it is in the electromagnetic searches now going on for SETI (Niedermeyer, Lopes da Silva, 2005).

**Discussion**

At first glance it would appear that this project should prove to be simpler than some of the existing SETI experiments, since we would only be looking at an EEG frequency range of 0.1-70 Hz initially, with the possibility of expanding to 3,000 Hz if necessary. This search range can be compared to Project Phoenix with a microwave search between 1.2-3 GHZ, the SETI project examining the frequency of the 21 cm (1,420 MHz) line of neutral hydrogen, or SERENDIP examining 168 million narrow (0.6 Hz) channels in a 100 MHz band centered at 1.42 GHz. The problem is that the EEG frequencies and amplitudes can vary widely from cycle to cycle, whereas the SETI frequencies are very constant as regards both Hz and amplitude.

However, on the other side of the coin, it is this variability of these EEG frequencies and amplitudes that may assist us in our search, in that any combinations of the two, that either repeat or fall outside the usual range of EEG activities, either in a repetitive or random fashion which has never been observed before, could point us in a SETI-type direction. The other advantage which a biological nonlocal search would have, is that we could *respond* to any unusual signals of this nature by stimulating either the MEAs or the human subjects with the same type signals that were received, and know that they will probably be transmitted anywhere in the universe *instantaneously*!

One of the first objections to this proposal probably has to do with why, if we are supposedly being bombarded with all this nonlocal communication from what could be an enormous number of extraterrestrial sources, we do not notice it in some fashion. And, you can apply this same reasoning to just the citizens of earth where, if this effect really exists, one would think we should be inundated with this type of superluminal communication, based on a population of $6 \times 10^9$! The reason why we are not, can be explained by looking at the research data previously discussed. This also answers the question as to how one can differentiate between nonlocal communication from subjects here on earth vs that from extraterrestrials. You will recall that when the human subjects were stimulated by either plain flashes of light or by checkerboard patterns, the stimulated subject showed a VEP, especially as regards amplitude of the waves, several orders of magnitude *larger* than the amplitudes in the non-stimulated subject's EEG. In addition, in all these research studies, no type of transference of conscious subjective experience was ever observed, which would be the main criteria revealing that actual communication had taken place between the two parties. It evidently takes a very unusual stimulus to elicit such a response, and is what prompted me to address this problem by advocating other types of stimuli (Thaheld, 2003; 2006b). This means then that *most* of the nonlocal communication is probably lost within the normal brain wave or EEG noise, otherwise we would be incapable of functioning normally and would overload when confronted with such an influx of data.

This is not to say that we do not sporadically (and at times so swift either ignore it or doubt that something really happened) experience these nonlocal communication events. There are numerous anecdotal instances where twins or relatives knew that something was happening or had happened to a loved one but, this usually occurred when an unusually strong but brief stimulus was sent and received, like in an accident or disaster. It may well be that we experience these events especially during periods of sleeping or dreaming, when we are unconscious, and the usually prevalent EEG waking noise levels are reduced. Experiments have been proposed to test this theory. Thus it may well be that consciousness serves to protect us while we are awake, by putting most of this nonlocal communication into an unreachable noise zone. If you have ever been to a rock concert, you know what it is like trying to carry on a conversation with the person next to you. Having researched this several decades ago, one only has to look at the resulting EEG to get an idea of these elevated brain noise levels over and above the decibel levels!

One only has to look at our inability to perceive any of the total electromagnetic spectrum beyond the narrow frequency range for vision, without having to resort to different pieces of specialized equipment. Even in the case of vision, it is interesting to note that only about 10% of the total photons incident upon the cornea, are ever transduced and amplified by the retina, and received as information by the visual cortex.

**Conclusion**

Whether one believes in this experimental approach or not, it has the advantage that it would be very simple and cheap to carry out. The proposed SETI project can either go forward simultaneously on several different fronts or in just one specific area at a time and, while it would be nice to be able to prove the existence of biological nonlocality and CSC here on earth before embarking on this SETI approach, we may be forced to look outside the confines of the earth initially, as outlined. Nevertheless this search can be conducted in the following fashion but, not necessarily in the sequence shown.

1. MEAs containing human neurons can be subjected to specifically chosen random laser pulses of varying Hz either individually or in pairs, and then left to themselves for a period of hours or days while any electrical responses are being recorded. This collected data can then be run through specially developed algorithms for computer-assisted analysis. The alternative is to not initially stimulate these basins with a laser and just see what signals of an unusual nature are generated.

2. Human subjects are subjected to specifically chosen stimuli of an unusual repetitive or random nature as regards both Hz and amplitude, and AEEG recordings can be taken over a selected period of hours or days, with particular emphasis during those periods of sleeping when very slow delta waves are prevalent, with once again the data being run through special algorithms for computer-assisted analysis. This can also be performed without any initial stimulus and just analyzing any signals which are received.

3. Development of SETI-type algorithm software so that any electroencephalographers or neurophysiologists around the world could also perform similar SETI computer-assisted analysis of their regular patients in addition to their usual analysis of their EEGs.

4. It would appear that the critical elements in any EEG SETI-type search will involve an empirical blend consisting of the proper algorithms, filters, viewing windows, amplification and either compression or expansion of the derived data.

**Acknowledgements**

H., Thiemann, W., Brack, A., Greenberg, J.M., 2002. Amino acids from ultraviolet irradiation of interstellar ice analogues. Nature 416, 401-406.

Ćirković, M.M., 2004. The temporal aspect of the Drake equation and SETI. Astrobiology 4, 225-231. astro-ph/0306186.

Ćirković, M.M., 2006. Macroengineering in the galactic context. In. *Macro-Engineering: A Challenge for the Future.* ed. V. Badescu, R.B. Cathcart, R.D. Schuiling. Springer, New York. astro-ph/0606102.

Crawford, I., 2000. Where are they? Searching for extraterrestrials. Scientific American. July, 39-43.

Dirac, P.A.M., 1963. The evolution of the physicist's picture of nature. Scientific American. May, 208.

Drake, F., Sobel, D., 1991. *Is anyone out there: The search for extraterrestrial intelligence.* Delacorte, New York.

Duane, T.D., Behrendt, T., 1965. Extrasensory electroencephalographic induction between identical twins. Science 150, 367.

Eberhard, P.H., 1978. Bell's theorem and the different concepts of nonlocality. Il Nuovo Cimento 46B, 392-419.

Einstein, A., Podolsky, B., Rosen, N., 1935. Can quantum mechanical description of physical reality be considered complete? Phys. Rev. 47, 777-780.

Everett, H., 1954. "Relative state" formulation of quantum mechanics. Rev. Mod. Phys. 29, 454-462.

Feynman, R., 1982. Simulating physics with computers. Int. J. Theor. Phys. 21, 467-488.

Gaidos, E., Selsis, F., 2006. From protoplanets to protolife: the emergence and maintenance of life. astro-ph/0602008.

Gato-Rivera, B., 2005. A solution to the Fermi paradox: The solar system, a part of a galactic hypercivilization? physics/0512062 v2.

Gilliam, F., Kuzniecky, R., Faught, E., 1999. Ambulatory EEG monitoring. J. Clin. Neurophysiology 16, 111-115.

Grinberg-Zylberbaum, G., Delaflor, M., Attie, L., Goswami, A., 1994. The Einstein-Podolsky-Rosen paradox in the brain: the transferred potential. Phys. Essays 7, 422-429.
36